\documentstyle[eqsecnum,aps,manuscript]{revtex}
\def\be{\begin{equation}}
\def\ee{\end{equation}}
\begin{document}
\draft
\title{LOW--ENERGY FORMULAS FOR NEUTRINO MASSES  
WITH $\tan\beta$-DEPENDENT HIERARCHY}  
\author{M.K.Parida}
\address{High Energy Physics Group,\\
International Centre for Theoretical Physics, I--34100 Trieste, Italy\\ 
and\\ 
Physics Department, North--Eastern Hill University, P.O.Box 21,Laitumkhrah\\ 
Shillong 793 003,India}
\author{N.Nimai Singh}
\address{Physics Department, Gauhati University, Guwahati 781 014, India}
\date{October 13,1997}
\maketitle
\begin{abstract}
Using radiative corrections and seesaw mechanism, we derive analytic 
formulas for neutrino masses at low energies in SUSY unified theories
exhibiting a  new hierarchial relation,for the first time, among 
them. The new hierarchy is found to be quite significant especially for 
smaller values of $\tan\beta$.
\end{abstract}
\narrowtext
Unification of gauge and Yukawa couplings in supersymmetric grand 
unified theories and their predictions have received  considerable 
attention over the last decades [1-4]. In addition to solving 
the well-known gauge hierarchy problem, SUSY unified theories based 
upon $N=1$ supergravity, or originating from superstrings, 
provide an elegant method of unification with gravity. Independent 
of these, the problem of neutrino masses has intrigued experimental 
as well as theoretical physicists since more than sixty years. 
If a unified theory, 
based upon supergravity or superstrings, is to be accepted as the 
ultimate fundamental theory of nature, fermion masses including those 
of neutrinos have to emerge as necessary predictions of such a theory. 
The simplest and the most elegant method of obtaining neutrino masses 
is through seesaw mechanism [5] leading to the Majorana masses of 
left--handed neutrinos:
\be m_{\nu_i}=C_{\nu_i}{m^2_i\over M_N}\;,\;i=1,2,3\eqnum{1}\ee 
where  $\nu_i\,(i=1,2,3)=(\nu_e,\nu_\mu,\nu_\tau)$ and 
$m_i(i=1,2,3)=(m_u,m_c,m_{\rm top})$. With $C_{\nu_i}\simeq 1$, 
the canonical seesaw formula is valid at the scale $(\mu=M_X)$ 
where, due to spontaneous symmetry breaking, a higher gauge 
symmetry such as SUSY SO(10), or one of its intermediate gauge 
groups, breaks down to the SUSY standard model and the right--handed 
Majorana neutrinos acquire mass $M_N$ ( assumed degenerate). 
But the limits on neutrino masses extending over wide range of 
values, are estimated either from cosmology or experimental 
measurements like tritium $\beta$-decay, neutrinoless double 
$\beta$-decay, solar and atmospheric neutrino fluxes and neutrino 
oscillations at LSND [6]. Therefore, it is of utmost importance and 
considerable interest that accurate predictions of the seesaw 
mechanism are made available at low energies. For the first time, 
Bludmann, Kennedy and Langacker (BKL) [7] found that $C_{\nu_i}$'s 
deviate drastically from unity due to radiative corrections. 
Very significant refinements using semi-analytic and numerical methods, 
including renormalisation of Yukawa couplings and neutrino-mass 
operator, have been carried out in SUSY [8] and nonSUSY SO(10) [8-9]. 
In this work, we derive low-energy  formulas for neutrino masses, 
analytically, in SUSY unified theories demonstrating,for the first time,
the existence of a new hierarchial relation among them.
We also estimate the impact of such a new hierarchy on 
neutrino-mass ratio predictions for interesting regions of 
$\tan\beta$ corresponding to $t-b-\tau$ and $b-\tau$ 
Yukawa unification models.In the small $\tan\beta$ region  the mass 
ratios are found to be very significantly dependent upon $\tan\beta$.
\par 
The derivation of analytic formulas is carried out using the 
renornalisation group equations (RGEs) at one--loop level for 
gauge couplings $(g_i(t)\,,\,i=1,2,3)$, Yukawa couplings 
$(h_i(t))$ [8,10--12], neutrino mass operator $(K(t))$ [8] and vacuum 
expectation value (VEV) of $u$-type Higgs doublet $(V_u(t))$ [11] 
in SUSY standard model embedded in GUTs like SO(10) without 
[3,10--12] or with an intermediate gauge symmetry [4],
\be {dh_f(t)\over dt}=Oh_f(t)\;,\;f=u,c,{\rm top},b,\tau\eqnum{2}\ee
\be Oh_{\rm top}={h_{\rm top}\over 16\pi^2}\left[\;6h_{\rm top}^
2+h^2_b-\sum^3_{i=1}C_ig^2_i\;\right]\eqnum{3}\ee
\be Oh_b={h_b\over 16\pi^2}\left[\;6h_b^2+h^2_\tau+
h_{\rm top}^2-\sum^3_{i=1}C^{\prime}_ig^2_i\;\right]\eqnum{4}\ee
\be Oh_\tau={h_\tau\over 16\pi^2}
\left[\;4h_\tau^2+3h^2_b-\sum^3_{i=1}C''_ig^2_i\;\right]\eqnum{5}\ee
\be Oh_{u,c}={h_{u,c}\over 16\pi^2}\left[\;3h_{\rm top}^2-
\sum^3_{i=1}C_ig^2_i\;\right]\eqnum{6}\ee
\[ C_i=(\;{13\over 15}\;,\;3\;,\;{16\over 3}\;)\] 
\[ C'_i=(\;{7\over 15}\;,\;3\;,\;{16\over 3}\;)\] 
\[ C''_i=(\;{9\over 5}\;,\;3\;,\;0\;)\nonumber\]
\[ {dg_i\over dt}={b_ig_i^3\over 16\pi^2}\;,\;i=1,2,3\]
\be b_i=(\;{33\over 5}\;,\;1\;,\;-3\;)\eqnum{7}\ee
\be {dK\over dt}={K\over 16\pi^2}\left[\;6h_{\rm top}^2+
2h_\tau^2-{6\over 5}g_1^2-6g_2^2\;\right]\eqnum{8}\ee
 
In eqs.(2)--(8), $i=1,2,3$ correspond to the gauge groups $U(1)_Y$, 
$SU(2)_L$, and $SU(3)_C$ of the SUSY standard model assumed
to hold for $\mu\geq m_{\rm top}=M_{SUSY}$. The variable $t=\ln\mu$ is 
the natural logarithm of the scale parameter $\mu$. 
We include scale $(t=\ln\mu)$ dependence in the Yukawa couplings 
[8,10-12] as well as the vacuum expectation value [11] in the 
definition of the quark masses,
\be m_{u,c,{\rm top}}(t)={h_{u,c,{\rm top}}(t)V_u(t)\over\sqrt 2}
\eqnum{9}\ee 
Using eq.(9) and evolving $K$ from the higher mass scale $\mu=M_X$ 
to $\mu=\mu_0=m_{\rm top}$, the renormalized coefficients in the 
seesaw formula are,
\be m_{\nu_i}(t_0)=C_{\nu_i}(t_0){m^2_i(t_0)\over 
M_N}\;,\;i=1,2,3\eqnum{10}\ee
\be C_{\nu_i}(t_0)=\left[\;{h_{{\rm top},c,u}(t_x)\over 
h_{{\rm top},c,u}(t_0)}\;\right]^2
\left[\;{K(t_0)\over K(t_x)}\;\right]\eqnum{11}\ee 
where $t_0=\ln m_{\rm top}$ and $t_x=\ln M_X$. It is to be noted that, the 
running VEV[11] cancels out from in eq.(11) yielding the same renormalization
as noted in ref.[8]. Now using RGEs (2)--(8), we obtain in a straight forward
manner,
\be C_{\nu_\tau}(t_0)=R(t_0)\,e^{(6I_{\rm top}+2I_b-2I_\tau)}\eqnum{12}\ee
\be C_{\nu_{\mu,e}}(t_0)=R(t_0)\,e^{-2I_\tau}
\eqnum{13}\ee
\[ R(t_0)=\sum^3_{i=1}\left[\;{\alpha_i(t_x)\over 
\alpha_i(t_0)}\;\right]^{d_i\over b_i}\;,\nonumber\]
\be d_i=\left(\;{-4\over 15}\;,\;{0}\;,\;{-16\over 
3}\;\right)\eqnum{14}\ee
\be I_f={1\over 16\pi^2}\int^{(t_x)}_{(t_0)}h_f^2(t)\,dt\;,\;f={\rm top}
,b,\tau .\eqnum{15}\ee 
In evolving the masses and gauge couplings to scales below 
$\mu_0=m_{\rm top}$, we ignore negligible contribution to 
$C_{\nu_i}$ between the scales $M_Z$ and $m_{\rm top}$. 
Below $\mu=M_Z\,(t_z=\ln M_Z)$, the renormalisation of the 
quark masses, $m_c$ and $m_u$, occur due to the presence of the 
gauge symmetry $SU(3)_C\times U(1)_{em}$ leading to well-known 
QCD--QED rescaling factors [11],
\be\eta_u\simeq {m_u(0)\over m_u(t_0)}\;,\;\;\eta_c\simeq 
{m_c(t_c)\over m_c(t_0)}.\eqnum{16}\ee
We then obtain
\be m_{\nu_\mu}(t_c)\simeq C_{\nu_\mu}(t_c){m_c^2(t_c)\over 
M_N}\eqnum{17}\ee
\be m_{\nu_e}(0)\simeq C_{\nu_e}(0){m_u^2(0)\over M_N}\eqnum{18}\ee 
where
\be C_{\nu_\mu}(t_c)={R(t_0)\over \eta^2_c}\,e^{-2I_\tau}
\eqnum{19}\ee
\be C_{\nu_e}(0)={R(t_0)\over 
\eta^2_u}\,e^{-2I_\tau}\eqnum{20}\ee 
In eqs.(16)--(20), $t_c=\ln m_c$ and $t_1=\ln\mu_1=0$ for 
$\mu_1=1$GeV. The quantities in the R.H.S. of (12) have negligible 
contributions due to renormalisation effects 
carried down to lower energies, $\mu=\mu_1=1$GeV $(t=t_1=0)$ although 
those in (13) change significantly. Combining eqs.(10)--(20), 
we obtain the low energy formulas for the left--handed Majorana 
neutrinos of three generations,
\be m_{\nu_\tau}=R\,e^{(6I_{\rm top}+2I_b-2I_\tau)}{m^2_{\rm top}\over
M_N}\eqnum{21}\ee
\be m_{\nu_\mu}={R\over \eta^2_c}\,e^{ -2I_\tau}{m^2_c\over 
M_N}\eqnum{22}\ee
\be m_{\nu_e}={R\over \eta^2_u}\,e^{ -2I_\tau}{m^2_u\over 
M_N}\eqnum{23}\ee 
where $R=R(t_0)$ as defined in (14) and the masses occuring on the R.H.S. 
are the low--energy values. The formulas (21)--(23) predict
low energy values of neutrino masses, analytically ,in a large class of 
SUSY unified theories including SO(10) which can be compared with the experimental
values.If the right-handed neutrino masses are nondegenerate,these formulas can be
easily generalised by replacing $M_N$ by $M_{N_ i}$ for the ith generation in
eqs.(21)-(23).But independent of any assumed value of the degenerate right-handed
neutrino mass ,the formulas predict a new hierarchial relation, 
\FL\[ m_{\nu_\tau}:m_{\nu_\mu}:m_{\nu_e}::m^2_{\rm 
top}\exp(6I_{\rm top}+2I_b)\nonumber\]
\be:(m^2_c/\eta^2_c):(m^2_u/\eta^2_u)\eqnum{24}\ee 
Using the experimentally observed values of the quark masses 
$m_u,\,m_c$ and $m_{\rm top}=174$GeV, the CERN--LEP measurements, 
and $\eta_u$ and $\eta_c$ [12], we have evaluated the central values 
and uncertainties of $C_{\nu_i}\,(i=e,\mu,\tau)$, for different 
values of $\tan\beta={V_u\over V_d}$, the ratio of VEVs of the 
$u$--type and $d$--type Higgs doublets. The details of such 
analyses and their aapplications to different SO(10) models will be reported
elsewhere. It is quite clear that the low-energy predictions on neutrino masses
depend crucially on the nature and values of Yukawa couplings between $m_{\rm top}$
and $M_X$.For example,if the Yukawa couplings increase at faster rate in certain 
GUTs,which may occur due to appearance of new degrees of freedom at higher mass
scales ,the new contributions to integrals over squares of Yukawa 
couplings , are likely to predict neutrino masses different from conventional SUSY
SO(10).However,here we confine to numerical results on  the relative hierarchy 
between $m_{\nu_\tau}$ and $m_{\nu_\mu}$ or 
$m_{\nu_e}$ in conventional SUSY GUTs, which doe not depend upon the arbitrariness
on $M_N$ ,
\be {m_{\nu_\tau}\over m_{\nu_\mu}}={m^2_{\rm top}\over 
m^2_c}\eta^2_c\,e^{(6I_{\rm top}+2I_b)}\eqnum{25}\ee
\be {m_{\nu_\tau}\over m_{\nu_e}}={m^2_{\rm top}\over m^2
_u}\eta^2_u\,e^{(6I_{\rm top}+2I_b)}\eqnum{26}\ee 
It may be recognised that the second and third factors in 
the R.H.S. of (25)--(26) are due to radiative corrections from 
$M_X$ down to low energies, corresponding to the ratios,
\[ {C_{\nu_\tau}\over C_{\nu_\mu}}
=\eta^2_c\,e^{(6I_{\rm top}+2I_b)}\nonumber\]
\be {C_{\nu_\tau}\over C_{\nu_e}}=\eta^2_u\,e^{(6I_{\rm 
top}+2I_b)}\eqnum{27}\ee
\par 
For larger values of $\tan\beta$ corresponding to $t-b-\tau$ 
Yukawa unification in 
supergrand desert models, the Yukawa couplings at the GUT scale 
and hence the values of $I_b$ ,$I_\tau$, and $I_{\rm top}$ change slowly with 
$\tan\beta$. The value of 
$\exp\left[\,6I_{\rm top},\right]$ almost remains constant 
giving rise to $C_{\nu_\tau}/C_{\nu_e}\simeq 15-20$ and 
$m_{\nu_\tau}/m_{\nu_e}\simeq (2-3)\times10^{10}$ for 
$\tan\beta=10-59$.As compared to the BKL value[7] of $m_{\nu_\tau}/
m_{nu_e}\simeq  9.2\times10^9$,the enhancement in the hierarchy in this 
region is only 2-2.5.
But as $\tan\beta$ decreases and approaches 
the value of 1.7, as in $b-\tau$ unification models with a 
super grand desert,$h_{\rm top}(M_X)$ and hence
$I_{\rm top}$ increase with the decrease of $\tan\beta$ while 
$h_b(M_X)$ ,$h_\tau$ , $I_b$ and $I_\tau$ are negligible. Thus,we find that,for
smaller values of $\tan\beta\simeq 1.8$,$I_{\rm top}$ = 0.4 and  the ratio
$(m_{\nu_\tau}/m_{\nu_e})$ has a 
substantial contribution due to the rise of the top quark Yukawa 
coupling near the unification scale which leads to the central 
value  of $C_{\nu_\tau}/C_{\nu_e}\simeq 83.6$ and 
$m_{\nu_\tau}/m_{\nu_e}\simeq 1.1\times10^{11}$.These may be compared
the results of BKL[7] corresponding to $C_{\nu_\tau}/C_{\nu_e}\simeq 7.6$
and $m_{\nu_\tau}/m_{\nu_e}\simeq 9.2\times10^{9}$.Whereas in the large 
$\tan\beta$ region,our enhancement factor is only 2-3,for smaller
$\tan\beta$=1.8,the enhancement factor ia as large as $\simeq 10$.  
Similar features are also noted in intermediate scale SUSY SO(10) model 
corresponding to $M_X=10^{13}$GeV where the intermediate gauge symmetry 
breaks down to the standard gauge group giving large mass to $M_N$ [4]. 
When the ratio  $m_{\nu_\tau}/m_{\nu_\mu}$ is  examined as a function of 
$\tan\beta$, similar type of enhancements are noted. For example, in the 
supergrand desert type models $(M_X\simeq 2\times 10^{16}$GeV), 
$C_{\nu_\tau}/C_{\nu_\mu}\simeq 65$ and 
$m_{\nu_\tau}/m_{\nu_\mu}\simeq 1.2\times 10^{6}$ 
for  $\tan\beta=1.8$, exhibiting an enhancement of hierarchy by a 
factor $\simeq 15$ over the BKL value[7] of $m_{\nu_\tau}/m_{\nu_\mu}\simeq
8.1\times10^{4}$GeV.The details of evaluation of neutrino masses and mass 
ratios including uncertainties,as a function of intermediate scales and 
$\tan\beta$ would be reported elsewhere.
\par 
 In summary we have obtained analytic formulas for left--handed Majorana--
neutrino masses at low energies demonstrating, for the first time ,the 
existence of a  new hierarchy between the neutrino masses which depends 
quite significantly upon $\tan\beta$  in the region of  smaller values
of this parameter. These results hold true in a large class of models
including SO(10). We suggest that these formulas be used while estimating 
neutrino mass predictions in unified theories with a degenerate right--handed 
Majorana neutrino mass. The formulas can be easily generalised 
for nondegenerate right-handed neutrino masses.
\par
After this work was completed,the contents of ref.[13] was brought to our notice
where  the renormalization for $m_{\nu_\tau}$  has been discussed,but to prove the
new hierarchy low energy formulas for all the three neutrino masses ,as
demonstrated here,are essential.
 
\acknowledgments 
One of us (M.K.P) thanks Professor R.N.Mohapatra  
for useful discussions and to Dr.K.S.Babu for  useful comments and
bringing  ref.[13] to our notice.The authors acknowledge support from the 
Department of Science and Technology, New Delhi through the 
research project No. SP/S2/K--09/91.


\begin{references}
\bibitem{1} For a review see W.de Boer, Prog.Part.Nucl.Phys. 
{\bf 33}, 201 (1994).
\bibitem{2} P.Langacker and M.Luo, Phys.Rev.D {\bf 44}, 
817 (1991); U.Amaldi, W.de Boer and H.Furstenau, 
Phys.Lett.B {\bf 260}, 447 (1991).
\bibitem{3} S.Dimopoulos, L.J.Hall and S.Raby, Phys.Rev.Lett. 
{\bf 68}, 1984 (1992); Phys.Rev.D {\bf 45}, 4192 (1992); 
G.Anderson, S.Dimopoulos, L.J.Hall, S.Raby and G.Starkman, 
Phys.Rev.D {\bf 49}, 3660 (1994); K.S.Babu and R.N.Mohapatra, 
Phys.Rev.Lett. {\bf 74} 2418 (1994); ibid. {\bf 70}, 2845 (1993).
\bibitem{4} Dae--Gyu Lee and R.N.Mohapatra, Phys.Rev.D 
{\bf 52}, 4125 (1995); M.Bando, J.Sato and T.Takashashi, Phys.Rev.D 
{\bf 52}, 3076 (1995);M.K.Parida, ICTP Report IC/96/33, Phys.Rev.D 
( to be published ).
\bibitem{5} M.Gell--Mann, P.Ramond and R.Slansky, in Supergravity, 
Proceeding of the Workshop, Stony Brook, New York 1997, eds. 
P.Van Nieuwenhuizen and D.Freedman 
( North Holland, Amsterdam, 1980 ); T.Yanagida, in Proceedings of 
the Workshop on Unified Theory and Baryon Number of the 
Universe,  eds. O.Sawada and A.Sugamoto, 
(KEK, Tsukuba ) 95(1979); R.N.Mohapatra and G.Senjanovic, Phys.Rev.Lett. 
{\bf 44}, 912 (1980); Phys.Rev.D {\bf 23}, 165 (1981).
\bibitem{6} For a recent review see J.Brunner, CERN Report, 
CERN--PPI/97 - 38 (1997) (unpublished).
\bibitem{7} S.A.Bludmann, D.C.Kennedy and P.G.Langacker, 
Nucl.Phys. {\bf B374}, 373 (1992).
\bibitem{8} K.S.Babu, C.N.Leung and J.Pantaleone, Phys.Lett.B 
{\bf 319}, 191 (1993).
\bibitem{9} M.K.Parida and M.Rani, 
Phys.Lett.B {\bf 377}, 89 (1996).
\bibitem{10} S.G.Naculich, Phys.Rev.D {\bf 48}, 5293 (1993); 
N.G.Deshpande and E.Keith, Phys.Rev.D {\bf 50}, 3513 (1994).
\bibitem{11} H.Arason, D.J.Castano, E.J.Piard and P.Ramond, Phys.Rev.D 
{\bf 47}, 232 (1993); 
H.Arason, D.J.Castano, B.Keszthelyi, S.Mikaelian, E.J.Piard, P.Ramond 
and B.D.Wright, Phys.Rev.D {\bf 46}, 3945 (1992).
\bibitem{12} V.Barger, M.S.Berger and P.Ohmann,Phys.Rev.D {\bf 47}, 
1093 (1993).
\bibitem{13} K.S.Babu,Q.Y.Liu, and A.Yu.Smirnov,hep-ph/9707457.
\end{references}
\end{document}